\documentclass[10pt,a4paper,twoside]{article}
\usepackage{amssymb}
\usepackage{amsmath}
\usepackage{fancyhdr}
\usepackage{amsthm}
\usepackage{rotate}
\usepackage{graphicx}
\usepackage[T1]{fontenc}

\usepackage{afterpage}  %
\theoremstyle{plain}
\newtheorem{theorem}{Theorem}[section]

\theoremstyle{definition}
\newtheorem{definition}[theorem]{Definition}

\theoremstyle{remark}

\usepackage[linesnumbered,ruled,vlined]{algorithm2e}
\SetKwInput{KwInput}{Input}                
\SetKwInput{KwOutput}{Output}              
\usepackage{multirow, makecell}

\begin{document}

\afterpage{\rhead[]{\thepage} \chead[\small  MohammadSadegh Mohagheghi and Khayyam Salehi      
]{\small  ML-based algorithm for Probabilistic Bisimulation in MDPs} \lhead[\thepage]{} }                  

\begin{center}
\vspace*{2pt}
{\Large \textbf{Improving Probabilistic Bisimulation for MDPs}}\\[3mm]
{\Large\textbf{ Using Machine Learning }}\\[30pt]
{\large \textsf{\emph{Mohammadsadegh Mohagheghi  $^{\star} $\footnote{$ ^{\star} $Corresponding author (E-mail: mohagheghi@vru.ac.ir)} and Khayyam Salehi}}}
\\[30pt]
\end{center}
\begin{abstract}
The utilization of model checking has been suggested as a formal verification technique for analyzing critical systems. However, the primary challenge in applying to complex systems is state space explosion problem. To address this issue, bisimulation minimization has emerged as a prominent method for reducing the number of states in a labeled transition system, aiming to overcome the difficulties associated with the state space explosion problem. In the case of systems exhibiting stochastic behaviors, probabilistic bisimulation is employed to minimize a given model, obtaining its equivalent form with fewer states. Recently, various techniques have been introduced to decrease the time complexity of the iterative methods used to compute probabilistic bisimulation for stochastic systems that display nondeterministic behaviors. In this paper, we propose a new technique to partition the state space of a given probabilistic model to its bisimulation classes. This technique uses the PRISM program of a given model and constructs some small versions of the model to train a classifier. It then applies machine learning classification techniques to approximate the related partition. The resulting partition is used as an initial one for the standard bisimulation technique in order to reduce the running time of the method. The experimental results show that the approach can decrease significantly the running time compared to state-of-the-art tools.

\vskip 3mm




\end{abstract}

\section*{\centerline{ 1. Introduction}}
In today's world, computers are everywhere, and when they malfunction, the effects can be profound. Furthermore, proving the accuracy of computer systems is important since some safety features that fail could endanger human life. One mistake in the launch of a rocket, for instance, might have a negative impact on the entire operation~\cite{clarke2018introduction}.

Testing is a promising approach to ensure the correctness of a system. However, it is unable to cover all possible scenarios and verify the system's correctness entirely~\cite{baier2008principles}. In contrast, formal methods utilize mathematical techniques to determine if a system would function correctly under all potential circumstances. There are two widely used formal methods: theorem proving and model checking. The former employs mathematical proofs to establish the program properties of the system, often requiring expert involvement. Conversely, model checking automatically verifies that the entire system behavior satisfies the desired properties~\cite{baier2008principles}. This paper will focus on adopting the model checking approach.

Model checking is an approach for formally verifying qualitative or quantitative properties of computer systems. It involves using a Kripke structure or labeled transition system to represent the underlying system and employing temporal logic or automata to specify the desired properties. By utilizing software tools, the proposed model is automatically checked to determine if the specified properties can be guaranteed. Given the stochastic nature of many computer systems, probabilistic model checking is available to verify properties of such systems. Markov decision processes (MDPs) and Discrete-time Markov chains (DTMCs) are two extensions of transition systems used for modeling stochastic computer systems~\cite{katoen2016probabilistic}. DTMCs are suitable for modeling fully probabilistic systems, while MDPs can capture both stochastic and non-deterministic behaviors of computer systems~\cite{katoen2016probabilistic}. MDPs often involve uncertainty and randomness in their decision-making processes. MDPs provide a powerful framework for studying such scenarios and finding optimal strategies or policies to achieve desired objectives under uncertain conditions. This makes them valuable tools in various fields, including artificial intelligence, control theory, operations research, and robotics, among others.

The primary obstacle in model checking is the state space explosion problem, wherein the size of models grows exponentially as the number of components increases. This limitation restricts the explicit representation of large models~\cite{clarke2018introduction, baier2008principles, katoen2016probabilistic}. To address this challenge, various techniques have been developed over the past few decades. Symbolic model representation~\cite{mcmillan1993symbolic,parker2003implementation}, compositional verification~\cite{clarke1989compositional, feng2010compositional}, statistical model checking~\cite{agha2018survey, legay2019statistical, legay2015statistical}, and reduction techniques~\cite{kwiatkowska2006symmetry, ciesinski2008reduction, hansen2011partial} are among the key approaches proposed to tackle this problem. These techniques are widely utilized in model checking tools to alleviate the impact of the state space explosion problem.

One of the techniques used for model reduction is bisimulation minimization, which establishes an equivalence class on the state space of the model~\cite{baier2020probabilistic, salehi2022automated}. States within each equivalence class, known as bisimilar states, share the same set of properties. By applying a bisimulation relation, states within a class can be collapsed into a single state, resulting in a reduced model that is equivalent to the original one. Importantly, the reduced model preserves the same set of properties, allowing a model checker to utilize it as a substitute for the original model~\cite{baier2008principles}.

The literature defines various types of bisimulation depending on the class of transition systems and the properties being considered. One commonly discussed type is strong bisimulation, where two states, $ s $ and $ t $, are considered bisimilar if, for every successor state of $ s $, there exists at least one bisimilar successor state of $ t $, and vice versa~\cite{baier2008principles, groote2018efficient}. Another type is weak bisimulation, which disregards silent transitions and defines bisimilar states based on a path that includes some silent moves along with a move having the same action~\cite{baier2020probabilistic}. In this paper, our focus is on strong bisimulation, and we propose a machine learning technique to reduce the computational time of iterative algorithms used to compute this particular version of the bisimulation relation for probabilistic systems. Further information about other classes of bisimulation and their associated algorithms can be found in~\cite{cattani2002decision}.

In previous works, numerous techniques have been proposed for computing probabilistic bisimulation. The initial works on defining bisimulation for probabilistic automata and Markov Decision Processes (MDPs) can be traced back to~\cite{larsen1989bisimulation, segala1995modeling}. Definition of both strong and weak bisimulation for probabilistic systems incorporating non-determinism, along with their associated algorithms, was first introduced in~\cite{stoelinga2002verification}. These works have contributed significantly to the development of techniques for analyzing and verifying probabilistic systems using bisimulation. 

To improve the performance of the standard algorithms for computing probabilistic bisimulation in MDPs, we propose a novel approach. The proposed approach uses machine learning techniques to directly compute bisimilar equivalence classes. The computed partition can be used as the initial one for an iterative partition refinement algorithm. One of the benefits of such an approach is its capability to extend to other types of bisimulation or transition systems. To the best of our knowledge, no previous work has used machine learning for classifying bisimilar blocks of states. In summary, the main contributions of our work are as follows:
\begin{itemize}
\item
We use machine learning to classify the state space of a model to its bisimilar blocks. Our technique uses several small versions of a given model for the training step.
\item 
Because of the different number of bisimilar classes, we use the concept of superblock to gather several similar blocks. In this way, the number of superclasses is the same among different models of a probabilistic program.
\end{itemize}

The structure of the paper is as follows. In Section 2, we review some preliminary definitions of MDPs and probabilistic bisimulation and the standard algorithm for computing a probabilistic bisimulation partition. In Section 3, we describe the proposed approach for using machine learning to compute bisimilar classes of a given MDP model. Section 4 provides the experimental results running on several classes of the standard benchmark models. Finally, Section 5 concludes the paper and introduces some future work.

\section*{\centerline{ 2. Preliminaries}}\setcounter{section}{2}\setcounter{theorem}{0}
\label{sec:preliminary}
In the context of a finite set $ S $, a distribution $ \mu $ over $ S $ is a function $ \mu:S\rightarrow[0,1] $ that assigns non-negative values to each element of $ S $ such that $\sum_{s\in S}\mu(s) = 1$. In other words, for every $ s\in S $, the value $ \mu(s) $ represents the probability associated with $ s $. The set $ S $ is considered as the state space, and each member $ s\in S $ is referred to as a state.

The set of all distributions over $ S $ is denoted by $ D(S) $. It encompasses all possible functions that satisfy the conditions of being a distribution over the set $ S $. Furthermore, given a subset $ T\subseteq S $ and a distribution $ \mu $, the accumulated distribution over $ T $, denoted as $ \mu[T] $, is defined as the sum of the probabilities or weights assigned to the states within $ T $. Mathematically, it is defined as $\mu[T] = \sum_{s\in T}\mu(s)$.

A partition $\mathcal{B} $ of a set $ S $ consists of non-empty and disjoint subsets, forming equivalence blocks, which cover the entire set $ S $. An equivalence relation $ R $ is defined based on the blocks, where two states are considered equivalent if they belong to the same block. Formally: $ s\;R\;t $ if and only if $ \exists B_i\in\mathcal{B}$ s.t. $s,t\in B_i $. 

The set of equivalence classes of $ R  $ on $ S $ is denoted as $ S/R $. For a subset $ T  $ of $ S $, $ T/R $ represents the set of states related to at least one state in $ T $, that is, $ T\subseteq S $,  $ T/R=\{s\in S|\ \exists t\in T:sRt\} $. A partition $ \mathcal{B}_1 $ is considered finer than $ \mathcal{B}_2 $ if every block in $ \mathcal{B}_1 $ is a subset of a block in $ \mathcal{B}_2 $. Additionally, the equivalence relation $ R $ on distributions over $ S $ can be extended by comparing accumulated distributions within the equivalence blocks defined by the partition, formally: $ \mu R\nu $ if and only if $ \mu[C]=\nu[C] $ for every block $ C\in S/R $. That is, the accumulated distribution $ \mu $ of $ C $ is the same as $ \nu $.

\begin{definition}[Markov Decision Process (MDP)]
	 An MDP $ M $ represents as a tuple $ (S,s_0,Act,\delta,G) $ where $ S $ is a finite set of states, $ s_0\in S $ is an initial state, $ Act $ is a set of finite actions, $ \delta:S\times Act\rightarrow\mathcal{D}(S) $ is a (partial) probabilistic transition function, which maps a state and an action to a distribution of states, and $ G $ is the subset of states representing the set of goal states.
\end{definition}

	MDPs are widely used as mathematical models to represent and analyze systems that exhibit both non-deterministic and probabilistic behavior. The number of states in the MDP is denoted by $ \left|S\right| $, and the number of actions available is represented by$  \left|Act\right| $. The set of actions enabled in each state $ s $ is denoted by $ Act(s) $. In other words, in state $ s $, we can select an action $\alpha\in Act\left(s\right) $. 
	
	MDP $ M $ works as follows. It starts by selecting an initial state $ s_0 $. Once MDP $ M $ is in a particular state $ s $, a nondeterministic choice between the enabled actions needs to be resolved. Suppose action $ \alpha\ \in\ Act(s) $ is non-deterministically chosen. Then, according to the induced distribution $ \mu=\delta(s,\alpha) $, the next state $ s' $ is probabilistically specified. To resolve non-deterministic choices of an MDP, the notion of policies (also known as adversaries) is utilized. A policy is a (deterministic) mapping that associates each state $ s\in S $ with a specific enabled action $ \alpha\in Act(s) $. 
	
	Reachability properties of probabilistic systems are determined as the probability of achieving a set of states of the model. For MDPs, the properties can be determined as the extremal (maximal or minimal) probability of reaching a goal state $ G $ over all possible policies. In bounded reachabilities, the number of steps that can be taken is restricted to a predetermined bound. More comprehensive details on probabilistic model checking and the specific iterative methods used for computing reachability properties can be found in~\cite{baier2008principles}.
	
\begin{definition}[Probabilistic Bisimulation]
	A probabilistic (strong) bisimulation $ R\subseteq S\times S $ is an equivalence for $ M $ if and only if for each pair of states $ s,t\in S $, if $ s $ has a relation to $t \ (sRt)$  then for every action $ \alpha\in Act(s) $ there exists an action $ \beta\in Act(t) $ such that $ \delta\left(s,\alpha\right)\ R\ \delta\left(t,\beta\right) $. In this case, the probability of going each block is the same for both actions. The name of actions is irrelevant to characterized the bisimilarity of two states; whereas in probabilistic automata, actions names should be taken into account.
\end{definition}
Two states $ s,t\in S $ are probabilistically bisimilar if and only if there exists a probabilistic bisimulation $ R $ such that $ sRt $. In the literature, probabilistic bisimulation is characterized in terms of a goal set of states $ G $; that is, we have either $ s,t\in G $ or $ s,t\in S\setminus G $ for any pair of bisimilar states $ s $ and $ t $.

The key feature of probabilistic bisimulation is that for any pair of bisimilar states $ s,t\in S $, the same set of bounded and unbounded reachability properties are satisfied in both states~\cite{baier2008principles}. Consequently, a reduced bisimilar (especially smaller) MDP can be created by exchanging all bisimilar states of any block $ B_i\in\mathcal{B} $ of original MDP $ M $ by one state.

\subsection{The Standard Algorithm for Computing a Probabilistic Bisimulation}
\begin{figure*}[ht!]
	\centering
		\includegraphics[scale=0.4]{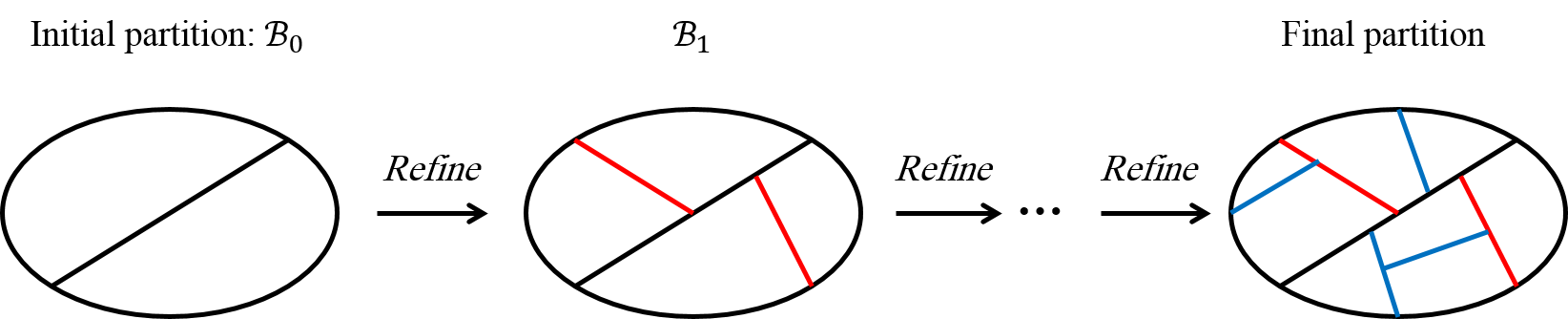}
		\caption{Successive partition refinement procedure}
		\label{fig:refine}
\end{figure*}

Partition refinement is a widely applicable algorithm for computing a bisimulation relation in various types of transition systems. The algorithm begins with an initial partition and proceeds iteratively by refining the partitions through the splitting of certain blocks into smaller, more refined blocks. The iterations continue until a fixed point is reached, meaning that no further splitting of blocks is possible (Figure~\ref{fig:refine}).

During each iteration, a splitter block is chosen to divide some predecessor blocks into smaller and finer ones. The specific method of splitting a block depends on the definition of bisimulation tailored to the underlying transition system being analyzed. Algorithm~\ref{alg:general_bisimulation} outlines the steps involved in this approach~\cite{baier2008principles}. 
\begin{algorithm}[ht!]
	\caption{Partition refinement algorithm}
	\label{alg:general_bisimulation}	
	\KwInput{An MDP model M}
	\KwOutput{bisimulation partition $\mathcal{B}$}
	Initialize $\mathcal{B}$ to a first partition\;
	
	\While{{there is a splitter  for $\mathcal{B}$}}
	{
		Choose a splitter $C$ for $\mathcal{B}$\;
		$\mathcal{B} := \mathit{Refine}(\mathcal{B}, C)$\;
	}
	\Return{ $\mathcal{B}$}\;
\end{algorithm}

In probabilistic bisimulation, the refinement procedure for partitioning blocks takes into account the probabilities associated with reaching a splitter block $C$. This procedure involves splitting a block $ B_i $ from the current partition $ \mathcal{B} $ into multiple subblocks $ B_{i,1},B_{i,2},\cdots,B_{i,k} $ based on the following conditions:
\begin{enumerate}
	\item $\cup_{1\leq j \leq k}B_{i,j} = B_i$,
	\item $B_{i,j}\cap B_{i,l} = \emptyset$ for $1\leq j < l \leq k$,
	\item for each $1\leq j \leq k$ and every two states $s,t\in B_{i,j}$, it holds that for each action $\alpha\in Act(s)$ there is an action $\beta\in Act(t)$ where $\delta(s,\alpha)[C] = \delta(t,\beta)[C]$.
\end{enumerate}

By utilizing an efficient data structure, the time complexity of the \textit{Refine} method in Algorithm~\ref{alg:general_bisimulation}~(Line 4) is in $ O(|M|+|S|\cdot|Act|\cdot\log{\left|Act\right|}\ ) $~\cite{groote2018efficient}. In the algorithm, a queue of blocks is used, where after refining each block, all computed subblocks, except the largest one, are added to the queue as potential splitters. This strategy ensures that each state is considered in some splitters for at most $ \log(|S|) $ times. Based on this approach, the overall time complexity of Algorithm~\ref{alg:general_bisimulation} for computing probabilistic bisimulation is in $ O(|M|\cdot\log{\left|S\right|}+|S|\cdot\log{\left|S\right|}\cdot|Act|\cdot\log{\left|Act\right|}) $.

There are various approaches to compute the initial partition $ \mathcal{B}\subseteq S\times S $. One possible method is to consider two blocks, $ G $ and $ S\setminus G $, as the initial partition and use $ G $ as the first splitter. Using a finer initial partition, Algorithm~\ref{alg:general_bisimulation} requires fewer iterations to reach the fixed point. In this paper, a novel heuristic is proposed for computing the initial partition. This heuristic incorporates a machine learning technique to approximate the relevant partition of the probabilistic bisimulation relation. Ideally, the approximated partition aligns with the final partition of the probabilistic bisimulation relation, yielding the best-case scenario.
\subsection{The PRISM Modeling Language}
\begin{figure*}[ht!]
	\centering
		\includegraphics[width=0.75\linewidth]{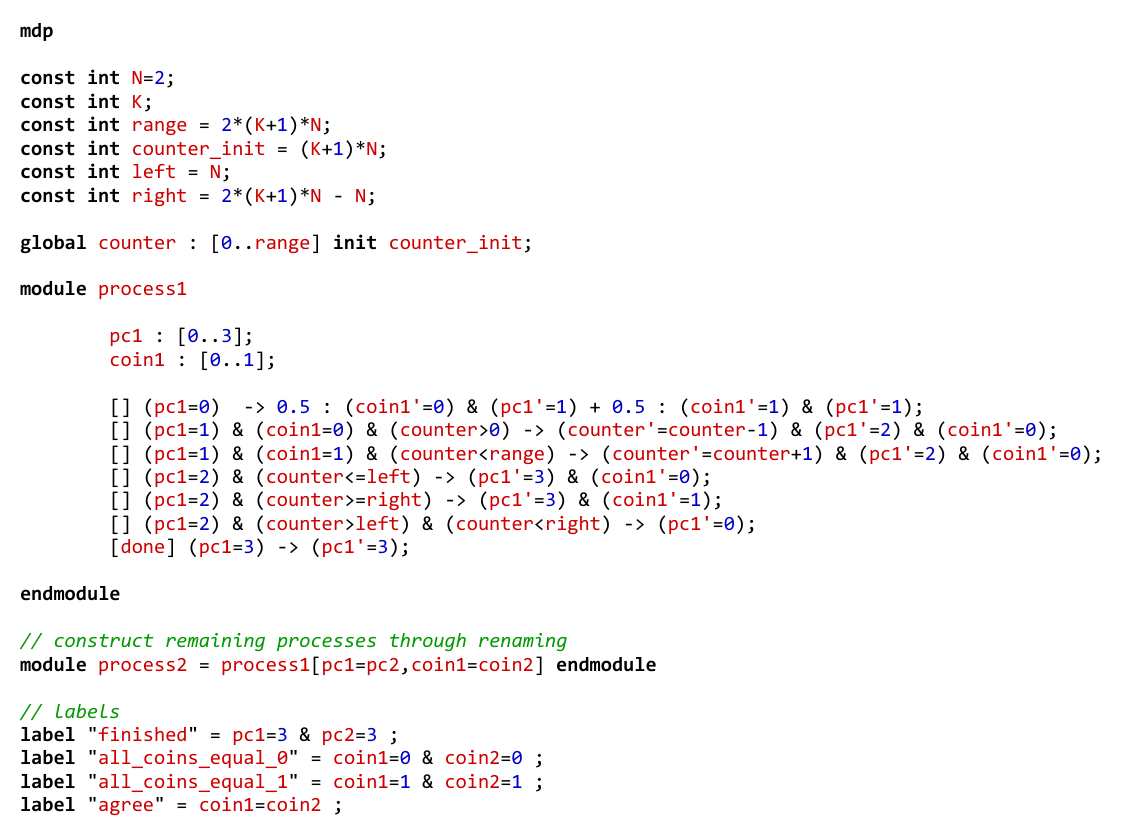}
		\caption{The PRISM code for the \textit{Coin} MDP model.}
		\label{fig:PRISM}
\end{figure*}

The standard approach in model checking is to use a high-level modeling language to propose a description of the underlying system. A model checker translates the proposed program to a transition system as the semantics of the model. PRISM programs~\cite{kwiatkowska2011prism} can be used for the case of probabilistic model checking. In this modeling language, each program contains one or more modules, while each module has several variables with a defined domain of values. Several guarded commands describe possible transitions of the model. A probabilistic model checker (such as PRISM~\cite{kwiatkowska2011prism} or STORM~\cite{hensel2022probabilistic}) parses a program to a related MDP or DTMC. An example of a PRISM program is proposed in Figure~\ref{fig:PRISM}.

It defines an MDP model with two modules \textit{process1} and \textit{process2} while the second module is a copy of the first one. The first module has two variables \textit{pc1} and \textit{coin1} with defined domain of values. Moreover, several constants with known values and a parameter constant ($ K $) are used in the definition. A global variable \textit{counter} is defined that its upper-bound is determined by $ K $. Thus, using different values for the parameter $ K $, we may have different models with different sizes. 

Any valid valuation for the set of model variables induces a state of its associated transition system. However, only the set of states that are reachable from the initial state are needed for model checking and are stored explicitly or implicitly. Formally, for the set of model variables, a state $ s_i\in S $ maps any of these variables to a value in its domain. For a set $ v_i,v_j,\cdots,v_k $ of state variables, we use $ \prod_{v_i,v_j,...,v_k}(s)\ $  as a projection function, which gets a list of the value of these variables in $ s $. For any subset $ R\subseteq S $, we define$  \prod_{v_i,v_j,...,v_k}(R)=\cup_{s\in R}\prod_{v_i,v_j,...,v_k}(s) $. For a model $ M $ of a given PRISM program, we use $ \mathit{Vars(M)} $ for the set of its variables (not constants) and use $ \mathit{Params(M)} $ for those variables where upper bounds are bounded by a parameter. We call such variables parametric. For the induced MDP $ M $ of Figure~\ref{fig:PRISM}., we have $ \mathit{Vars(M)}=\{counter,pc1,\ coin1,\ pc2,\ coin2\} $ and $ \mathit{Params(M)} =\left\{counter\right\} $. The MDP variable \textit{counter} is parametric because its upper bound is determined by the model parameter $ K $.

\section*{\centerline{ 3. The Proposed Approach}}\setcounter{section}{3}\setcounter{theorem}{0}\setcounter{subsection}{0}
In this section, we propose a novel heuristic for approximating the initial partition. The correctness of the approximated is checked by the partition refinement method (Algorithm~\ref{alg:general_bisimulation}). If the approximated partition requires more refinements, it can be used as a more precise initial partition which may result in faster convergence towards the fixed point. In other words, this can be considered as a preprocessing step of partition refinement method.

For the sake of simplicity, we assume that every program graph has only one parameter. It should be noted that even with this assumption, a variable with a parametric value domain may have several copies in the model definition and also in the induced MDP. The general scheme of our approach is proposed in Algorithm~\ref{alg:Init_Partition}.
\begin{algorithm} [ht!]
	\caption{Approximating Initial Partition}
	\label{alg:Init_Partition}	
		\KwInput{A PRISM program with known values as parameters}
		\KwOutput{An approximated partition $\mathcal{B}$ for the induced MDP}

		 Construct several sample models $M_{p_1},M_{p_2},...,M_{p_n},$ using smaller values for the model parameter (Section~\ref{sec:construct-sample-model})\;
		 For each sample model $M_{p_i}$, apply the probabilistic bisimulation algorithm and compute its equivalence partition $\mathcal{B}_{p_i}$
		(Algorithm~\ref{alg:general_bisimulation})\;
		 For each partition $\mathcal{B}_{p_i}$, compute the set $\{sp_1, sp_2, ..., sp_k \}$ of its superblocks (Section~\ref{sec:construct-superblock})\;
		 Let $\eta(s)$ denote the superblock that $s$ belongs to.
		(
		Definition~\ref{def:superblock})\;
		 Fix a classifier and use $\eta$ for training (Section~\ref{sec:training})\;
		 Use the trained classifier to predict which superblock each state of the underlying model belongs to (Section~\ref{sec:classify})\;
		 Split states of each superblock to their blocks according to their parameter values  (Section~\ref{sec:split-superblock})\;
		 \Return {the partition $\mathcal{B}$ including the computed blocks of step 7\;}
\end{algorithm}

In the following subsections, we explain each step in detail. Recall that the main purpose of our approach is to facilitate the computation of bisimilar blocks for a large model, where the running time may be an obstacle. Even in non-precise computed blocks, they can be used to reduce the main model to an abstract version to cope with memory limitations.

\subsection{Constructing sample models and computing probabilistic bisimulation}
\label{sec:construct-sample-model}
As the first step of our approach, we consider several sample models by using smaller values $ p_1,p_2,\cdots,p_n $ for the parameter of the given PRISM program (Line~1 of Algorithm~\ref{alg:Init_Partition}). Depending on the structure of the given  program, the parameters can be so small that result in some tiny models or they may be large enough to have the same structure as the given model. In the next section, we explain more about the values of parameters for several case studies. Although the precision of machine learning may increase by using more samples, in practice using two or three sample MDP models with several thousand states may be enough. In this case, we have at least ten thousand states as training samples that are considered enough in the machine learning approach. Furthermore, we compute probabilistic bisimulation and the equivalence blocks of each model by utilizing the standard algorithm for MDP models of each probabilistic program (Line~2 of Algorithm~\ref{alg:Init_Partition}).

\subsection{Computing superblocks}\label{sec:Sprbl}
\label{sec:construct-superblock}
The main idea of our approach is to use a classifier to map each state of the underlying model to a block of the bisimulation equivalence relation. To do so, we consider each variable of an MDP as a feature of samples and each block as a class. Considering variable values of each state as its feature values, the classifier should determine which class (block) the state may belong to. An important challenge of using computed partitions of the sample models is that the number of blocks is different among different samples (versions) of an MDP model. In this case, a classifier is unable to map states to the correct classes. To cope with this challenge, we gather several blocks of a partition to a superblock. We define a superblock as a collection of several bisimulation blocks such that any state of a block has similar states in the other blocks where the variables are the same except the parametric variables (Line~3 of Algorithm~\ref{alg:Init_Partition}).
\begin{definition}
	\label{def:superblock}
	A superblock $sp$ of an MDP $M$ is the largest collection of blocks that for each pair of different blocks $B_i,B_j$, the following condition holds:
	\begin{displaymath}
		\forall s\in B_i \ \exists t\in B_j: \prod\limits_{\mathit{non-params}(M)}(s) = \prod\limits_{\mathit{non-params}(M)}(t).
	\end{displaymath}
\end{definition}

The intuition behind this definition is that by increasing the value of a parameter, we expect to have new blocks that are similar to some previous ones except for their parametric values that are higher than the others. In this case, the total number of subblocks does not change among different models of a PRISM program.

\begin{figure}[!t]
	\centering
	\begin{center}
		\includegraphics[width=0.6\columnwidth]{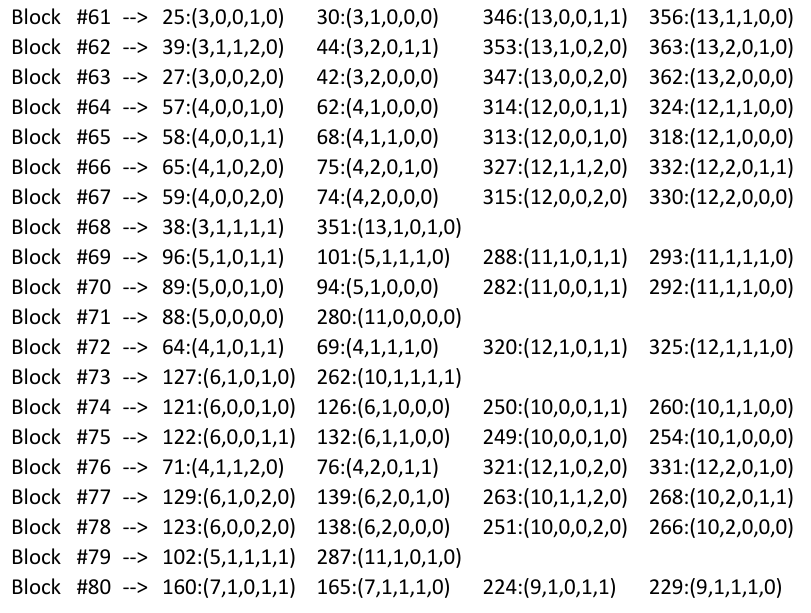}
		\caption{Some blocks of a Bisimulation Partition for the $\mathit{Coin}$ case study.}	
		\label{fig:coin-blocks}
	\end{center}
\end{figure}

For more clarification, consider Figure~\ref{fig:coin-blocks} that shows a list of some blocks of bisimilar states for the \textit{Coin} case study with $ K=3 $ as its parameter value. For each block, its number as the order that it is computed and the list of its states including state number and its feature values are reported. As an example, the 73’rd block contains two states: $ s_{127} $ and $ s_{262} $. For this case, a superblock contains the 68’th, 73’nd and 79’th blocks because the states of these blocks are of the form $ (x,1,0,1,0)$ and $(y,1,1,1,1) $ where $ x $ and $ y $ are the parametric variables.
To use superblocks for a classification process, we define $ \eta $ as mapping from states to superblocks (Line~4 of Algorithm~\ref{alg:Init_Partition}). For each state $ s\in S $ of a sample model, $ \eta(s) $ determines its corresponding superblock:
\begin{displaymath}
	\eta(s) = sp_i\ \emph{iff}\ \ \exists B\in sp_i,\ \ s\in B.
\end{displaymath}

\subsection{Training step}
\label{sec:training}
A classifier uses a set of superblocks for a training step. A Support vector machine (SVM) is used to accomplish this step. The purpose of this step is to prepare the classifier to predict classes of new state space. Hence, we use $ \eta $ as a mapping from state space to superblocks (Line~5 of Algorithm~\ref{alg:Init_Partition}). For a given PRISM program, we consider its variables as model features. For any parametric variable, we add the difference of the variable value and its domain upper bound (maximum value for the variable) as an additional feature of the model. For each state of a model, its features are considered as inputs to  $\eta $. This mapping is used to label each state for the training step.
\subsection{Classifying states into superblocks}\label{sec:classify}
For any state of a new model, the classifier determines its related superblock. The precision of a classifier for detecting the correct superclass depends on the structure of the models and associated PRISM programs. Because the training models and the given model have the same structure and only differ in their parameters, we expect to have promising results in most cases.

To improve the precision of our approach, we partition the state space of the given model to several subclasses according to its features. A subclass is assigned according to non-parametric variables of the PRISM program and possible values of these variables (reachable from the initial state). We apply these partitions for both training samples and the given model. For each subclass, a classifier is used to predict related superblocks of its states. In our approach, we first separate state space into several subclasses and then use a support vector machine classifier to improve the precision of classification (Line~6 of Algorithm~\ref{alg:Init_Partition}). 

\subsection{Splitting superblocks to bisimilar classes}
\label{sec:split-superblock}
As the final step of our approach, the states of each superclass should map to the correct bisimilar blocks. According to our definition of superblocks, a relation among parametric variables of bisimilar states of training sample models determines the possible values for the parameters of bisimilar states. For the example of Subsection~\ref{sec:construct-superblock}, the bisimilar states are as the form tuples $ (x,1,0,1,0) $ and $ (y,1,1,1,1) $ where $x+y\ =\ 16 $. For the parametric variable of this sample, we have $ counter\ =\ 16 $. For a given model with a known value for the counter parameter, our approach splits states of this superblock to bisimilar blocks where $ x\ +\ y\ =\ counter $ holds (Line~7 of Algorithm~\ref{alg:Init_Partition}). It is noteworthy that the proposed approach is an initialization step for partition refinement algorithm. The soundness of the approach is as the following.
\paragraph{Soundness.} To ensure the soundness of the approach proposed in this section, we consider several cases, where the initial partition may contain non-exact bisimilar states:
\begin{itemize}
	\item Case 1: A block $B'$ includes two or more bisimilar blocks $B_{j1},B_{j2},...$ where each $B_{jk}$ is a block of the correct bisimilar partition. This case happens in the standard bisimulation methods where a coarse relation is considered as an initial partition and the bisimulation algorithm terminates with a set of bisimilar blocks. 
	
	\item Case 2: A block $B''$ is proposed in the initial partition where it is a mere subset of an exact bisimilar block. In this case, at least one state $s$ exists that is bisimilar with the states of $B''$, but the method drops it in another block. For such initial partitions, the bisimulation algorithm results in a finer relation than the correct bisimilar one. Although such partition is not the minimized equivalence relation on the state space, it is sound and satisfies the same properties as a minimized model does~\cite{baier2008principles}.
	
	\item Case 3: A block contains some but not all states of two or more blocks. A standard bisimulation method will eventually divides the states of such block to several blocks of Case 2. In fact, some splitters will be used for this division. This leads to split some other blocks to finer ones of Case 2. Finally, the algorithm terminates where all blocks are either of Case 2 or the correct blocks.
\end{itemize}

\section*{\centerline{ 4. Experimental Results}}\setcounter{section}{4}\setcounter{theorem}{0}
In order to demonstrate the effectiveness and scalability of the proposed approach, we consider five classes of standard Markov decision process models. These MDP models serve as representative examples that cover a wide range of scenarios and characteristics. These classes include \textit{Coin}, \textit{Wlan}, \textit{Firewire}, \textit{Zeroconf}, and \textit{Brp} case studies from the PRISM benchmark suit~\cite{kwiatkowska2011prism}. All of them are parametric and are used to compare our machine learning-based approach. More details about these case studies are available at~\cite{kwiatkowska2011prism}.

We propose some information on the selected models in Table~\ref{table:models-training}. For each model, we report the parameter name in the third column while fixing another parameter (which is shown in the first column).
We have implemented our proposed approach as an extension to the PRISM model checker using PRISM 4.7, which is currently the last version. We implemented the proposed algorithm on a machine running Ubuntu 20.04 LTS with Intel(R) Core (TM) i7 CPU Q720@1.6GHz with 8GB of memory.
\begin{table*}
	\centering
	\footnotesize
	\caption{PRISM MDP models for training}\label{table:models-training}
	\resizebox{\textwidth}{!}{
	\begin{tabular}{cccc}
		\hline
		Model &  Number of  & Parameter & Number of states\\
		Name  &  variables  & names     &  for training   \\
		\hline
		\textit{Coin}($ N=4 $) & 9 & \textit{counter} & 76,032  \\
		\textit{Wlan}($ N= $5) & 13 & \textit{TRANS\_TIME\_MAX} & 2,794,536\\
		\textit{Firewire}($ delay=3 $) & 3 & \textit{deadline} & 710,924\\
		\textit{Zeroconf}($ N=900 $) & 22 & $ K $ & 577,128\\
		\textit{Brp}  & 18 & \textit{MAX} & 39,796\\ 
		\hline
	\end{tabular}}
\end{table*}

For the proposed approach and each computed partition, the set of superblocks are computed by using the proposed technique in the previous section. To simplify our approach, we gather all singular blocks (blocks with only one state) in one superblock. As the output of this step, we define $\eta$ as a mapping from states of each sample to the index of their corresponding superblock. This information is stored in some files (one file per sample model).

For the classification step, we develop our approach in Python. Our program reads the stored information of sample models including information of their states (variable values of each state) and the computed mapping. It separates the state space into several subclasses as explained in subsection 3.4. For each subclass, we apply SVM with its default parameters for classification. We first train the classifier for each subclass and then apply it to the states of a given model. In some cases, all states of a subclass are mapped to the same superblock and we need not to use a classifier for them. The number of states in training step on small sample models is shown in the last column in Table~\ref{table:models-training}.

To compare our implementation of the proposed methods for computing probabilistic bisimulation with the standard approaches, we consider PRISM~\cite{mohagheghi2023splitter}, STORM~\cite{hensel2022probabilistic}, and mCRL2~\cite{bunte2019mcrl2} as the well-known and state-of-the-art tools for computing probabilistic model checking for MDPs.

The experimental results are presented in Table~\ref{table:compare-approaches}. The number of states, actions, and transitions are shown in the third, fourth, and fifth columns, respectively. The running time for computing probabilistic bisimulation in PRISM, STORM, and mCRL2 as well as our proposed approach, ML-based, on the selected models are demonstrated in seconds. The Terms \textit{killed} and \textit{timeout} in Table~\ref{table:compare-approaches} refer to the out of memory error and the running time after one hour. We report the running time for computing the initial partition in sub-column \textit{init-part} and also the total running time after applying the partition refinement algorithm on initial partition in sub-column \textit{total}. The results show that the time for computing the initial partition of ML-based approach is approximately half of the total time for computing the partition refinement algorithm.
 
\begin{table}[ht!]
	\centering
	\caption{Performance comparison of computing bisimulation for the selected MDP with large values.}
	\label{table:compare-approaches}
	\resizebox{\textwidth}{!}{
	\begin{tabular}{llllllllll}
		\hline
		\multirow{2}{*}{\makecell{Model \\ name}} 
		& \multirow{2}{*}{Parameter} 
		& $|S|$ 
		& $| Act |$ 
		& $| Trns |$ 
		& \multirow{2}{*}{PRISM} 
		& \multirow{2}{*}{STORM} 
		& \multirow{2}{*}{mCRL2} 
		& ML-based \\
		\cline{9-10}
		&&$\times10^{-3}$&$\times10^{-3}$&$\times10^{-3}$&&&& init-part & total \\
		\hline
		
		\multirow{4}{*}{\makecell{\textit{Coin}\\ (N=5)}}   & K=30      & 2341    & 7832    & 9787     & 2.35  & 112              & 18.9  & 1.02     & 1.97  \\
		& K=50      & 3890    & 13016   & 16267    & 3.45  & 303              & 31.4  & 1.66     & 3.08  \\
		& K=70      & 5439    & 18200   & 22747    & 4.32  & 554              & 45.2  & 1.97     & 3.93  \\
		& K=100     & 7762    & 25976   & 32467    & 7.22  & 1178             & \textit{killed}     & 3.23     & 6.42  \\
		\hline
		\multirow{5}{*}{\makecell{\textit{Zeroconf}\\ (N=1500)}} & K=12      & 3753    & 6898    & 8467     & 9.33  &  \textit{killed}  & 22.7  & 1.01     & 2.76  \\
		& K=14      & 4426    & 8144    & 9988     & 12.57 &  \textit{killed} & 25.9  & 1.2      & 2.97  \\
		& K=16      & 5010    & 9223    & 11307    & 15.6  &  \textit{killed}   & 30.8  & 1.56     & 3.47  \\
		& K=18      & 5476    & 10085   & 12359    & 18    &  \textit{killed}   & 37    & 1.68     & 3.9   \\
		& K=20      & 5812    & 10711   & 13124    & 21.43 &  \textit{killed}  & 39.3  & 1.75     & 4.04  \\
		\hline
		\multirow{3}{*}{\makecell{\textit{Firewire}\\ (dl = 36)}} & ddl=3000  & 2238    & 3419    & 4059     & 2.08  & 2283             & 6.9   & 0.84     & 1.71  \\
		& ddl=10000 & 7670    & 11742   & 13936    & 10.73 & \textit{timeout} & 27    & 1.97     & 6.5   \\
		& ddl=15000 & 11550   & 17687   & 20991    & 13.02 & \textit{timeout} &  \textit{killed}     & 2.17     & 7.62  \\
		\hline
		\multirow{3}{*}{\makecell{\textit{Brp}\\ (N=400)}} & max=150   & 787     & 787     & 1087     & 0.41  & 2.9              & 2.4   & 0.35     & 0.88  \\
		& max=300   & 1567    & 1567    & 2167     & 0.81  & 12.5             & 5.1   & 0.6      & 1.37  \\
		& max=600   & 3127    & 3127    & 4327     & 2.19  & 51.4             & 10.6  & 1.45     & 3.61  \\
		\hline
		\multirow{2}{*}{\makecell{\textit{Wlan}\\ (N=6)}} & ttm=1000  & 8093    & 12543   & 17668    & 0.89  & 320              &\textit{killed}  & 0.71     & 1.61  \\
		& ttm=2500  & 12769   & 21925   & 27051    & 1.36  & 1900  & \textit{killed}  & 1.22     & 2.57 \\
		\hline
	\end{tabular}}
\end{table}
 
 For \textit{Coin} case study, the total running time of our machine learning-based approach outperforms the other tools. For example, when parameter $ K=100 $, our approach runs in 6.42 seconds, while PRISM runs in 7.22, STORM in 1178, and mCRL2 is \textit{killed} by out of memory error. 
 
 In case of \textit{Zeroconf}, STORM is \textit{killed} by out of memory error for the whole parameter values. Our approach reduced the total running time by 3 up to 5 times compared to PRISM, and 8 to 9 times compared to mCRL2. 
 
 For \textit{Firewire}, STORM has timeout in greater parameter values, while mCRL2 is killed by memory error. PRISM and ML-based approach run effectively on all models; whereas the running time of ML-based approach is half of the PRISM running time.
 
 On \textit{Brp} models, PRISM runs in the best running time compared to the other tools. Our approach takes more time than PRISM, but dominates STORM and mCRL2. In \textit{Wlan} models, mCRL2 is killed by out of memory error. The running time of STORM increases exponentially as the parameter value increases. Similar to \textit{Brp}, our approach takes more time compared to PRISM, but the time is approximately close to each other. In these two cases, using other classifiers rather than SVM may result in better performance. This is left as a future work.

\section*{\centerline{ 5. Conclusion}}\setcounter{section}{5}\setcounter{theorem}{0}
In this work, we have proposed a novel approach to improve the performance of the standard algorithms for computing probabilistic bisimulation for MDP models. The approach uses machine learning classification technique to even directly determine a bisimulation partition. Experimental results show that our approach outperform the other available tools. For the future work, we aim to extend the proposed technique to the other classes of transition systems such as probabilistic automata or discreate-time and continuous-time Markov chains. On can use other classifiers rather than SVM and compare their running time with the state-of-the-art tools. As another future work we plan to apply the proposed approach to analyze security protocols, especially anonymity protocols~\cite{shields2000protocol} such as dining cryptographers~\cite{noroozi2019secure}, single preference voting~\cite{salehi2019channel}, crowds~\cite{reiter1998crowds}, and TOR~\cite{reed1998anonymous}. 

\vskip 3mm

\noindent\textbf{Conflicts of Interest.} 
The authors declare that they have no conflicts of interest refarding the publication of this article.


%
%
%
%


\noindent   Mohammadsadegh Mohagheghi\\
Department of Computer Science,\\
        Vali-e-Asr University of Rafsanjan,\\
        Rafsanjan, I. R. Iran \\
         e-mail: mohagheghi@vru.ac.ir

\vskip 3mm

\noindent        Khayyam Salehi\\
Department of Computer Science,\\
        Shahrekord University,\\
        Shahrekord, I. R. Iran \\
         e-mail: kh.salehi@sku.ac.ir

\end{document}